\documentclass{emulateapj}
\usepackage{graphics}
\usepackage{psfig}
\usepackage{ulem}
\usepackage{longtable}

\def\b0540{{B0540$-$69}}
\def\j0537{{J0537$-$6910}}
\def\etal{{\it et~al.}}
\def\E{{\sl Einstein Observatory}}

\def\S{{\sl Swift}}
\def\SP{{\sl Swift}.}
\def\X{{\sl RXTE}}
\def\XP{{\sl RXTE}.}

\def\simlt{\lower.5ex\hbox{$\; \buildrel < \over \sim \;$}}

\begin{document}

\title{A New, Low Braking Index For the LMC Pulsar B0540$-$69}

\author{F.\ E.\ Marshall\altaffilmark{1},
L.\ Guillemot\altaffilmark{2,3},
A.\ K.\ Harding\altaffilmark{1},
P.\ Martin\altaffilmark{4},
D.\ A.\ Smith\altaffilmark{5},
}

\email{frank.marshall@nasa.gov}
\altaffiltext{1}{Astrophysics Science Division, NASA Goddard Space Flight Center, Greenbelt, MD 20771, USA}
\altaffiltext{2}{Laboratoire de Physique et Chimie de l'Environnement et de l'Espace
 (LPC2E), CNRS-Universit\'{e} d'Orl\'{e}ans, F-45071 Orl\'{e}ans, France}
\altaffiltext{3}{Station de radioastronomie de Nan\c{c}ay, Observatoire de Paris,
 CNRS / INSU F-18330 Nan\c{c}ay, France}
\altaffiltext{4}{Institut de Recherche en Astrophysique et Plan\'{e}tologie, UPS / CNRS, 
UMR5277, F-31028 Toulouse Cedex 4, France}
\altaffiltext{5}{Centre d'\'{E}tudes Nucl\'{e}aires de Bordeaux Gradignan, 
IN2P3 / CNRS, Universit\'{e} de Bordeaux 1, BP120, F-33175, Gradignan Cedex, France}
\shorttitle{Braking Index of B0540$-$69}
\shortauthors{Marshall et~al.}

\begin{abstract}

We report the results of a 16-month monitoring campaign using the \S\ satellite
of PSR \b0540, 
a young pulsar in the Large Magellanic Cloud.
Phase connection was maintained throughout the campaign so that a reliable
ephemeris could be determined, and the length of the campaign is adequate
to accurately determine the spin frequency $\nu$
and its first and second derivatives.
The braking index $n$ is $0.031 \pm 0.013$ (90\% confidence), a value much lower than
previously reported for \b0540\ and almost all other young pulsars.
We use data from the extensive monitoring campaign with \X\ to show that 
timing noise is unlikely to significantly affect the measurement.
This is the first measurement of the braking index in the pulsar's recently discovered
high spin-down state.
We discuss possible mechanisms for producing the low braking index.

\end{abstract}


\section{Introduction}
\label{introduction}

Rotation-powered pulsars gradually convert their rotational energy
to energetic particles and electromagnetic radiation,
which slowly reduces the rate of rotation.
A detailed understanding of this process, including the configuration of the pulsar's
magnetosphere and the physical mechanisms therein remains elusive.
The evolution of the spin frequency $\nu$ reveals information about the torques
acting on the neutron star and thus provides clues about the physical
processes at work in the neutron star
and its magnetosphere.
The gradual evolution of $\nu$ is typically described as a
Taylor series of $\nu$ and its time derivatives.
For many pulsar models $\dot\nu = -\kappa\nu^{n}$, where
$\kappa$ is related to properties of the neutron star and
$n$ is the ``braking index''.
For constant $\kappa$, $n = \ddot\nu \nu \dot\nu^{-2}$.
For the idealized case of magnetic dipole radiation from a spinning
neutron star in a vacuum, the braking index is 3
(e.g., Ostriker \& Gunn 1969).
Measurement of the braking index is usually difficult
because $\dot\nu$ changes very slowly for old pulsars,
and timing noise and sudden changes in the spin rate
(``glitches'') can be complications for young pulsars.

\b0540\ is a remarkable rotation-powered pulsar in the Large Magellanic Cloud (LMC).
Its spin-down luminosity ($-4\pi^2I\nu\dot\nu$)
is among the largest for all pulsars 
(ATNF Pulsar Catalog\footnote{http://http://www.atnf.csiro.au/reasearch/pulsar/psrcat},
Manchester \etal\ 2005).
It is by far the most luminous pulsar known in gamma-rays (Fermi LAT Collaboration, 2015).
\b0540\ has also been detected in the optical 
(Mignani \etal\ 2010 and references therein)
and radio (Manchester \etal\ 1993) bands.
The slowdown of \b0540\  
is relatively stable for a young pulsar, 
and it is one of only 9 young pulsars for which a braking index
has been reliably measured (Lyne \etal\ 2015; Ou \etal\ 2016; Archibald \etal\ 2016).
Except for J1640$-$4631, the braking indices for all 9 pulsars are smaller than that
expected for the simple magnetic dipole model.
We do not include the young pulsar J0537$-$6910, for which
Middleditch \etal\ (2006) presented evidence for $n=-1.5$ 
by noting the evolution of the minimum and maximum
values for $\dot\nu$ during intervals between the numerous glitches.
A much higher value for $n$ is seen during the phase-connected observations
between glitches. 
\b0540\ has been extensively studied 
since its discovery with the \E\ (Seward \etal\ 1984). 
Ferdman \etal\ (2015) used 15.8 years of data from 
the {\sl Rossi X-ray Timing Explorer} (\X)
to determine a braking index of $n = 2.129 \pm 0.012$ that
was consistent over the entire data span.

Marshall \etal\ (2015) recently discovered a transition to 
a new spin-down state for \b0540\ 
that had persisted for at least 1349 days.
Despite the extensive previous observations of \b0540\,
it appears that this new spin-down state
had never been seen before.
Rapid transitions between two spin-down states have
been seen for other pulsars (e.g., Lyne \etal\ 2010).
``Intermittent pulsars'' 
(Kramer \etal\ 2006; Young \etal\ 2013)
are extreme examples of such pulsars.
They transition from a radio-on state to radio-off state
with a simultaneous change in the spin-down rate.
More observations are needed to determine whether
\b0540\ should be considered an intermittent pulsar.

In this paper we present and discuss the timing analysis
of the continuing monitoring campaign
of \b0540\ with \SP\ 
The new measurements are combined with the earlier observations
with \S\ reported in Marshall \etal\ (2015)
and compared with results from the extensive monitoring
with \XP\
Details of the observations and
the results of the temporal analysis are reported in Section 2,
and interpretations are given in Section 3. 
Finally, Section 4 is a summary.

\section{Observations and Data Reduction}
\label{obs}
\b0540\ has been regularly observed with the X-Ray Telescope (XRT)
instrument (Burrows \etal\ 2005) on the {\sl Swift Gamma-Ray Burst Explorer}
(Gehrels \etal\ 2004) starting in Feb., 2015.
The XRT is a focusing X-ray telescope with a CCD detector with an effective
bandpass of $0.3-10$ keV. 
The cadence of the observations was chosen to maintain
phase information throughout the campaign.
We report on observations through June, 2016, and compare
our results to those from the earlier 12.5-year monitoring campaign
with the \XP\
Results from \X\ observations have been reported
by Zhang \etal\ (2001), Livingstone \etal\ (2005)
and Ferdman \etal\ (2015).

We followed the same data analysis procedures given
in Marshall \etal\ (2015).
X-ray events from each continuous observing interval
(a ``snapshot'')
were screened to maximize the signal from the pulsar,
including restricting the energy range to $0.5-8.0$ keV.
Arrival times were corrected to the solar system barycenter
using the {\sl Hubble Space Telescope} position of $\alpha = 05^{h}40^{m}11.202^{s}$,
$\delta = -69^{\circ}19^{\prime}54.17^{\prime\prime}$ (J2000.0)
(Mignani \etal\ 2010) and the JPL Planetary Ephemeris DE-200. 
The mean of the start and stop times of the snapshot was used
as the epoch of the snapshot.
Events were folded on multiple candidate pulsar periods, 
and a sine wave was fit to the best folded light curve.
The peak of the sine wave was used to determine the pulsar phase at the epoch.
The one $\sigma$ uncertainty in the phase was set to
$(\pi A)^{-1} (2N)^{-0.5}$, in which $A$ is the relative amplitude
of the sine wave and $N$ is the total number of counts.
The lengths of the snapshots ranged from 213 s
to 1796 s
because of the complexities of scheduling \S\ observations.
Typically the pulsar was not detected in the very short snapshots.
Table \ref{obs_log} is a log of all the snapshots used in this paper.

Uncertainties quoted in this paper are given at the 90\% confidence level 
unless otherwise noted.

\subsection{{\sl Swift}~XRT Timing Analysis}
\label{xrt}
The XRT started observations of \b0540\ on Feb. 17, 2015,
as part of a Target of Opportunity campaign. 
All observations were made using the Window Timing (WT) mode in which the central
200 CCD columns are continously read out, providing a time resolution
of 1.7 ms.
The XRT data were reduced with the standard software ({\scshape xrtpipeline}
v0.13.2) applying the default filtering and screening criteria 
({\scshape HEASoft 6.19}), using the latest updates to the XRT CALDB files
and to the clock correction file.
The phase and frequency were measured 
for each snapshot in which the pulsar was detected.

\begin{figure}
\includegraphics[angle=270.0,scale=0.6]{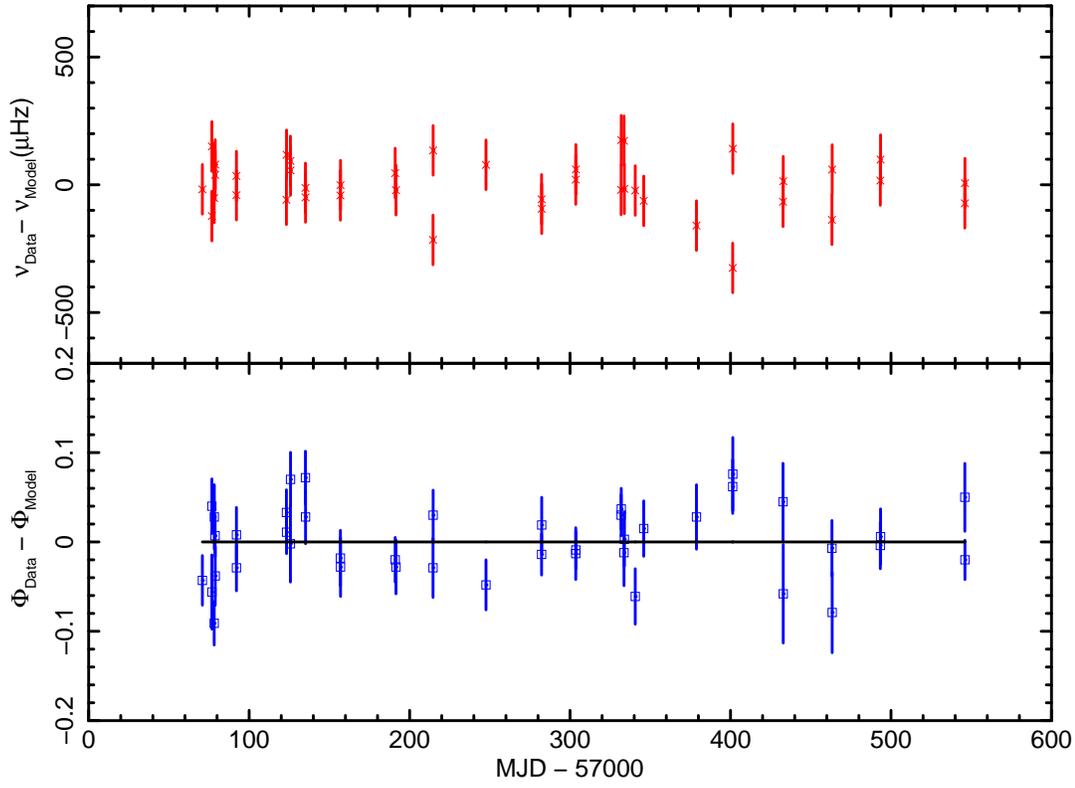}
\caption{The phase and frequency residuals for the XRT observations 
relative to the ephemeris model in Table \ref{ephemeris}.
The estimated one-$\sigma$ uncertainties are shown.
The solid line shows a residual of 0.0 for the phase.
\label{joint_residuals}}
\end{figure}

We used the following steps to determine the best-fit ephemeris 
(Table \ref{ephemeris}).
We started with the ephemeris reported by Marshall \etal\ (2015)
and computed the differences between the predicted frequencies for each snapshot
and the observed frequencies.
We then fit a linear model to the frequency differences
to produce an updated estimate for $\nu$ and $\dot\nu$.
We set $\ddot\nu$ to 0.0 because this process is not
sensitive to any plausible value for $\ddot\nu$.
Fitting the measured frequencies does not suffer
from the unavoidable integer-cycle phase ambiguity
that affects fitting phases, but it produces much larger
uncertainties for the ephemeris parameters.
Subsequent steps to determine the best ephemeris
used fits to the observed phases (rather than the frequencies).
We then did a grid search of possible values for
$\nu$ and $\dot\nu$ using the closely spaced
observations near MJD 57333, which contain 5 snapshots.
The grid was centered near the values obtained from
fitting the frequencies and covered more than twice
the 90\%-confidence error intervals for the parameters.
The grid search confirmed that there were no other
reasonable values for $\nu$ and $\dot\nu$.
We then gradually extended the observations being fit to earlier
and later times and found that the entire data set
could be fit after making small adjustments to
$\nu$ and $\dot\nu$ and adding a small $\ddot\nu$ term.
The result was a phase-connected ephemeris for the entire
time span of 16 months.
The final parameter values for the best-fit ephemeris
are given in Table \ref{ephemeris}, and the residuals
to values predicted by the best-fit ephemeris 
are shown in Figure \ref{joint_residuals}.
The value of 1.76 for the reduced $\chi^2$  for the fit
indicates a modest amount of unmodelled phase noise.
Because of this extra noise,
the uncertainties in Table \ref{ephemeris} are determined
by noting the change in the parameter corresponding
to an increase in $\chi^2$ of $4.77$ instead of the
usual $2.71$ for 90\% confidence.

Although the value for $\ddot\nu$ is unexpectedly low,
we are quite confident that we have determined the correct
ephemeris.
The procedure for modeling the observed frequencies is
not affected by any phase ambiguity, 
and uniqueness of the best-fit values for $\nu$ and $\dot\nu$
was confirmed with the grid search.
Substantially different values for $\ddot\nu$ are
constrained by the long time span of the observations.
For example, a $\ddot\nu$ of $3.7 \times 10^{-21}$ s$^{-3}$,
as reported by Ferdman \etal\ (2015),
produces a change in phase of $0.6 (\delta t/1 \times 10^7 s)^{3}$
where $\delta t$ is the time from the epoch time of the ephemeris.
If the correct value for $\ddot\nu$ is comparable (or larger than)
the value reported by Ferdman \etal\ (2015), then
as we expand the time interval being fit beyond $2 \times 10^7$ s
the phase residuals should start deviating from the
best-fit ephemeris. No such deviations are seen.
All 27 snapshots with
$\vert\delta t\vert > 1 \times 10^7$ s match the predicted
phase to within $\pm 0.1$.
This is compelling evidence that our best-fit
ephemeris is correct.

\subsection{Timing Noise in the \X\ and \S\ Eras}
\label{noise}

Pulsar timing noise is quasi-random changes in $\nu$
that can affect the measurement of $\ddot\nu$
in ways that are hard to predict.
The lack of any glitch in the 16 months of XRT observations
and the small phase residuals (Fig. \ref{joint_residuals})
to a third-order polynomial model
for this interval suggest that the timing
noise for the two spin-down states may be comparable.
We examined the \X\ data to determine the effect
of timing noise on $\ddot\nu$ as determined using
time spans of 16 months (the duration of the XRT observations).
We used 156 observations spaced over the span of 4560 days,
which was easily adequate to maintain phase information.
We divided the well sampled observations into nine
16-month intervals,
and computed the braking index for each of
intervals.
We then offset the intervals by 8 months and
repeated the analysis.
The usual Taylor series expansion including $\ddot\nu$
provided a good fit to the observed phases 
for all the intervals except those containing
the two small glitches (at MJD 51349 and 52925)
reported by Ferdman \etal\ (2015).
The glitches are readily apparent in the data
as gradually growing deviations from the predicted phase
over time scales of $\sim 1 \times 10^6$ s.
No such glitches are seen in the current campaign with \SP\
We eliminated the data affected by the glitches
from two of the intervals in each group.
A small spread in the braking index is found in each group.
The mean value for the first group is 2.115 with
a standard deviation of 0.018.
The mean for the second group is 2.131 with a standard
deviation of 0.050.
These results indicate timing noise has a relatively
small effect on the determination of the braking index
of \b0540\ when measured over 16-month intervals.
If the timing noise during the current campaign is
comparable to that seen during the \X\ campaign,
then it is very unlikely that the very low
braking index measured in the current campaign is
significantly affected by timing noise.

\section{Discussion}
\label{discussion}

\b0540\ has been extensively monitored by many
observatories in the more than 30 years since its discovery.
There have been numerous measurements of 
$\ddot\nu$ and thus the braking index,
first by Middleditch \etal\ (1987)
and most recently by Ferdman \etal\ (2015).
Boyd \etal\ (1995),
Deeter \etal\ (1999),
and Ferdman \etal\ (2015)
summarize the many observations
and discuss the measurement difficulties.
The values for the braking index range from a low
of $1.81 \pm 0.07$ (Zhang \etal\ 2001)
to a high of $2.74 \pm 0.10$ (\"{O}gelman \& Hasinger 1990)
for those measurements with small reported uncertainties.
The results can be affected by unseen glitches,
timing noise, inaccurate positions for the pulsar,
and instrumental timing errors, and
these effects are likely contributing to the
wide range of inconsistent values reported.
Using data from \X\ avoids or mitigates these problems.
The observations were sufficiently frequent to maintain
phase information and identify glitches,
an accurate pulsar position was known and could
be checked with the pulsar data, instrument time
tags were accurate, and the effect of timing
noise could be estimated (e.g., Section \ref{noise}).
The braking indices determined using the \X\ data by
Cusumano \etal\ (2003) ($2.125 \pm 0.001$),
Livingstone \etal\ (2005) ($2.140 \pm 0.009$), and
Ferdman \etal\ (2015) ($2.129 \pm 0.012$)
are very similar.
Further, Ferdman \etal\ (2015) report that a consistent
braking index is measured for the entire 15.8 years
covered.
Zhang \etal\ (2001) also used \X\ data, but their
low braking index may be affected by the inaccurate
position of the pulsar they used combined with the short data span
of only 289 days.
All these measurements of the braking index were for
times when \b0540\ was in its low spin-down state.
We conclude that the most reliable measurements
are from \X\, and that there is no compelling
evidence that the braking index has changed
since the pulsar was discovered until its transition
to a new spin-down state in 2011.

We report the first determination of the braking index
for the high spin-down state of B0540$-$69.
The measured value of $0.031 \pm 0.013$
is very much lower than and inconsistent with 
that of any of the previous measurements, 
which were all done during the low spin-down state.
This is compelling evidence that the transition
to the new spin-down state caused the change
in the braking index.

Such large changes in the braking index are unusual
for young pulsars. Previous reports of large
changes have been for pulsars
with high magnetic fields and the changes are induced
by large glitches.
Livingstone \etal\ (2011) reported a reduction in $n$
of $0.49 \pm 0.13$ for J1846$-$0258 after magnetar-like
behavior in 2006 including a glitch, an X-ray flux increase,
and spectral changes. Its magnetic field is $5 \times 10^{13}G$.
Archibald \etal\ (2015) confirmed this change
with a much longer time span of post-glitch observations.
Antonopoulou \etal\ (2015) reported a possible reduction
in $n$ by $\sim 15\%$ after glitches in 2004 and 2007
for J1119-6127, which has a surface dipole magnetic field of 
$4.1 \times 10^{13}G$.
A large change in $n$ was also reported in the high-magnetic-field
($7.4 \times 10^{13}G$) pulsar J1718-3718 (Manchester \& Hobbs 2011),
but the very large negative values for $n$ are difficult to interpret.
Archibald \etal\ (2015) suggest that the large magnetic field
is responsible for the change in braking index for these pulsars.
Since the magnetic fields for these pulsars are much larger than
the $6 \times 10^{12}G$ field for B0540$-$69,
and the state change for B0540$-$69 was not caused by a glitch,
these may be different phenomena.

The increase in the spin-down rate of \b0540
by 36\% suggests the possibility of a new
component contributing to the spin-down rate.
This situation can be modelled as
$\dot\nu = -(A\nu^{n1} + B\nu^{n2})$
(c.f., Equation 8 in Lyne \etal\ 2015)
in which $A$ and the braking index $n1$
apply to both the low and high spin-down states
and $B$ and $n2$ apply only to the high
spin-down state.
The values of $A$, $B$, and $n2$ can be solved
for using the values of $\dot\nu$
and the braking index measured for the low and high
spin-down states.
A braking index of $2.129$ for the low 
spin-down state leads to $n2 = -5.8$.
This very unusual value for the braking index 
indicates that the change in
spin-down state is probably not caused by the addition
of a new component with the old spin-down
mechanism(s) unchanged.

Braking indices have been measured for 9 young
pulsars, including B0540$-$69.
Measured braking indices for young pulsars range from $0.9 \pm 0.2$
for J1734$-$3333 (Espinoza \etal\ 2011)
to $3.15 \pm 0.03$ for J1640$-$4631 (Archibald \etal\ 2016).
The new value for the high spin-down state of \b0540
is substantially lower than and inconsistent with 
the values for these young pulsars.
Except for J1640$-$4631, all the measured braking indices 
are less than 3, the value for pure magnetic dipole
radiation.
A wide range of explanations have been suggested for the low measured
braking indices including additional emission
components and changing values for the pulsar's magnetic
dipole moment $M$, its moment of inertia $I$, or
the inclination angle $\alpha$ between the magnetic
and rotation axes. 

Michel \& Tucker (1969) showed that for a pulsar wind model
the braking index is 1 rather than 3, and several
authors suggested this as an explanation for braking indices
$< 3$.
For example, Lyne \etal\ (2015) estimated the fractional
contribution of wind and dipole components for the Crab
and Vela pulsars based on their measured braking indices.
Kou \etal\ (2016) predicted the braking index 
for \b0540 in its high spin-down state
using a model with both magnetic dipole radiation
and a pulsar wind. 
A value of 1.79 was predicted using the vacuum gap
model of Ruderman \& Sutherland (1975).
Other acceleration models predict similar values for $n$
(Table 2 in Kou \etal\ 2016), all inconsistent
with our measured value.

Secular changes in the characteristics of the pulsar,
such as $I$, $M$ or $\alpha$,
could also have a significant effect on the braking index
(e.g., Blandford \& Romani 1988).
Gourgouliatos \& Cumming (2015) investigated the evolution
of the magnetic field in the crust of the neutron star
and concluded that this could explain the low values of
$n$ seen for young pulsars.
Ho (2015) explained the low braking indices of pulsars
as the result of a growing surface magnetic field
after it is buried at the birth of the neutron star.
For both these models, a rapid change in the rate of evolution would also be 
required to explain the change in $n$ seen
in the transition to the high spin-down state of B0540$-$69. 
Lyne \etal\ (2013) measured a steady increase in the separation
of the main radio pulse and the interpulse in the Crab pulsar
and noted that this could be caused by an increase in $\alpha$.
Lyne \etal\ (2015) suggest
that this steady increase could be responsible for the
difference in the measured braking index for the Crab of 2.50
and the nominal value of 3.0.
If this is the correct interpretation for B0540$-$69,
then a value of $\kappa / \dot\kappa$ of $831$ years is needed
for the high spin-down state and $3859$ years for the 
low spin-down state. 
For $\kappa \propto M^2 sin^2\alpha I^{-1}$,
the corresponding characteristic time scales for
$I$, $M$, and $\alpha$
are a factor of $1.0$, $2.0$, and $1.7$ (for $\alpha = 50^{\circ}$,
Zhang \& Cheng 2000) lower than the value for $\kappa$ respectively.
These time scales only apply to the high spin-down state, and would
have to abruptly change at the transition to this state.

Menou \etal\ (2001) and Chen \& Li (2016) investigated pulsar models
with torque from a surrounding disk that with plausible assumptions
will produce braking indices of $-1$ and $\sim 0$ respectively. 
These models can produce
the small measured braking index of $0.031$ by changing the disk accretion rate, 
but the simplest two-component model described above produces a
change in $\dot\nu$ much larger than that observed.

\section{Summary}

We report the first measurement of the braking index
for the high spin-down state of B0540$-$69.
The measured value of $0.031 \pm 0.013$
is much lower than any of the numerous previous measurements,
all of which were for the low spin-down state of B0540$-$69.
Using the extensive observing campaign with \X\,,
we show that timing noise is unlikely to significantly affect
our measurement.
This remarkably low braking index is not consistent with
models that combine magnetic dipole radiation and wind
components.
Torque from a surrounding disk could produce the low measured value,
but the simplest model does not reproduce the observed change
in $\dot\nu$.
Secular changes on a time scale of $\sim 500$ years 
in a pulsar parameter, such as the inclination angle
or magnetic field strength,
might cause such a low braking index.

\medskip
This work made use of data supplied by the UK Swift Science Data Centre at
the University of Leicester and the High Energy Astrophysics Science
Archive Research Center, provided by NASA's Goddard Space Flight Center.

\begin{deluxetable}{llccr}
\tabletypesize{\footnotesize}
\tablecaption{Observing Log\label{obs_log}}
\tablewidth{0pt}
\tablehead{
\colhead{Satellite} &
\colhead{Obs. ID} &
\colhead{Date} &
\colhead{MJD} &
\colhead{$T_{exp}$} \\
\colhead{}  &
\colhead{}  &
\colhead{(1)}  &
\colhead{(1)}  &
\colhead{(2)}  \\
}
\startdata
\S\ & 00033603002 & 2015 Feb 17 & $57070.90$ & $1.2$ \\
\S\ & 00033603004 & 2015 Feb 23 & $57077.75$ & $1.1$ \\
\S\ & 00033603004 & 2015 Feb 23 & $57077.81$ & $1.0$ \\
\S\ & 00033603005 & 2015 Feb 25 & $57078.21$ & $1.3$ \\
\S\ & 00033603005 & 2015 Feb 25 & $57078.28$ & $0.5$ \\
\S\ & 00033603006 & 2015 Feb 25 & $57078.81$ & $1.6$ \\
\S\ & 00033603006 & 2015 Feb 25 & $57078.89$ & $0.8$ \\
\S\ & 00033603007 & 2015 Mar 11 & $57092.05$ & $0.8$ \\
\S\ & 00033603007 & 2015 Mar 11 & $57092.11$ & $1.3$ \\
\S\ & 00033603008 & 2015 Apr 11 & $57123.20$ & $1.2$ \\
\S\ & 00033603008 & 2015 Apr 11 & $57123.27$ & $1.1$ \\
\S\ & 00033603009 & 2015 Apr 13 & $57125.73$ & $1.2$ \\
\S\ & 00033603009 & 2015 Apr 13 & $57125.80$ & $1.0$ \\
\S\ & 00033603010 & 2015 Apr 23 & $57135.10$ & $1.0$ \\
\S\ & 00033603010 & 2015 Apr 23 & $57135.17$ & $1.6$ \\
\S\ & 00033603011 & 2015 May 14 & $57156.85$ & $1.0$ \\
\S\ & 00033603011 & 2015 May 14 & $57156.91$ & $1.4$ \\
\S\ & 00033603012 & 2015 Jun 18 & $57191.04$ & $1.3$ \\
\S\ & 00033603013 & 2015 Jun 18 & $57191.53$ & $0.7$ \\
\S\ & 00033603014 & 2015 Jul 11 & $57214.59$ & $0.7$ \\
\S\ & 00033603014 & 2015 Jul 11 & $57214.65$ & $0.9$ \\
\S\ & 00033603015 & 2015 Aug 13 & $57247.58$ & $1.1$ \\
\S\ & 00033603016 & 2015 Sep 17 & $57282.24$ & $1.2$ \\
\S\ & 00033603016 & 2015 Sep 17 & $57282.31$ & $0.7$ \\
\S\ & 00033603017 & 2015 Oct 08 & $57303.52$ & $1.1$ \\
\S\ & 00033603017 & 2015 Oct 08 & $57303.58$ & $1.2$ \\
\S\ & 00033603018 & 2015 Nov 05 & $57331.85$ & $1.0$ \\
\S\ & 00033603018 & 2015 Nov 05 & $57331.91$ & $1.1$ \\
\S\ & 00033603019 & 2015 Nov 07 & $57333.58$ & $0.5$ \\
\S\ & 00033603019 & 2015 Nov 07 & $57333.84$ & $1.2$ \\
\S\ & 00033603020 & 2015 Nov 14 & $57340.63$ & $0.8$ \\
\S\ & 00033603021 & 2015 Nov 19 & $57345.88$ & $0.8$ \\
\S\ & 00033603022 & 2015 Dec 22 & $57378.73$ & $0.6$ \\
\S\ & 00033603023 & 2016 Jan 14 & $57401.32$ & $0.9$ \\
\S\ & 00033603023 & 2016 Jan 14 & $57401.39$ & $0.8$ \\
\S\ & 00033603024 & 2016 Feb 14 & $57432.70$ & $1.2$ \\
\S\ & 00033603024 & 2016 Feb 14 & $57432.84$ & $0.9$ \\
\S\ & 00033603025 & 2016 Mar 16 & $57463.20$ & $0.5$ \\
\S\ & 00033603025 & 2016 Mar 16 & $57463.34$ & $0.5$ \\
\S\ & 00033603026 & 2016 Apr 15 & $57493.32$ & $0.5$ \\
\S\ & 00033603026 & 2016 Apr 15 & $57493.52$ & $0.4$ \\
\S\ & 00033603030 & 2016 Jun 06 & $57546.02$ & $0.9$ \\
\S\ & 00033603030 & 2016 Jun 06 & $57546.08$ & $1.0$ \\
\enddata
\tablecomments{
(1)~At the start of the snapshot;
(2)~Exposure time in ks
}
\end{deluxetable}

\begin{deluxetable}{lc}
\tabletypesize{\footnotesize}
\tablecaption{Ephemeris Parameters\label{ephemeris}}
\tablewidth{0pt}
\tablehead{
\colhead{Parameter} &
\colhead{2015 - 2016} \\
\colhead{}  &
\colhead{(1)} \\
}
\startdata
Epoch (Modified Julian Date)\dotfill& 57281.24147641701 TDB\\
Phase\dotfill& 0.000 (15)\\
$\nu$ (Hz)\dotfill& 19.6963233901 (22)\\
$\dot\nu$ ($10^{-10} s^{-2}$)\dotfill& -2.5286507 (15)\\
$\ddot\nu$ ($10^{-21} s^{-3}$)\dotfill& 0.104 (41)\\
\enddata
\tablecomments{
(1)~The numbers in parentheses are the 90\% confidence errors
quoted in the last digit.
}
\end{deluxetable}

\clearpage

\end{document}